\documentclass{article}
\usepackage{xcolor}
\usepackage{titlesec}
\usepackage{amsmath}
\usepackage{amsthm}
\usepackage{setspace}
\usepackage{float}
\usepackage{graphicx}
\usepackage{listings}
\usepackage{fancyvrb}
\usepackage{seqsplit}
\usepackage{amssymb}
\usepackage{subcaption}
\titleformat{\section}[block]{\normalfont\large\bfseries}{\thesection}{1em}{}
\title{On Algorithmic Cache Optimization}
\author{Neil Bhavikatti, Cherry Creek High School \\ Mentor: Julien Langou, CU Denver}
\date{}
\newtheorem{theorem}{Theorem}

\begin{document}
\maketitle

\begin{abstract}
\begin{doublespace}
    We study matrix-matrix multiplication of two matrices, $A$ and $B$, each of size $n \times n$. This operation results in a matrix $C$ of size $n\times n$. Our goal is to produce $C$ as efficiently as possible given a cache: a 1-D limited set of data values that we can work with to perform elementary operations (additions, multiplications, etc.). That is, we attempt to reuse the maximum amount of data from $A$, $B$ and $C$ during our computation (or equivalently, utilize data in the fast-access cache as often as possible). Firstly, we introduce the matrix-matrix multiplication algorithm. Secondly, we present a standard two-memory model to simulate the architecture of a computer, and we explain the LRU (Least Recently Used) Cache policy (which is standard in most computers). Thirdly, we introduce a basic model Cache Simulator, which possesses an $\mathcal{O}(M)$ time complexity (meaning we are limited to small $M$ values). Then we discuss and model the LFU (Least Frequently Used) Cache policy and the explicit control cache policy. Finally, we introduce the main result of this paper, the $\mathcal{O}(1)$ Cache Simulator, and use it to compare, experimentally, the savings of time, energy, and communication incurred from the ideal cache-efficient algorithm for matrix-matrix multiplication. The Cache Simulator simulates the amount of data movement that occurs between the main memory and the cache of the computer. One of the findings of this project is that, in some cases, there is a significant discrepancy in communication values between an LRU cache algorithm and explicit cache control. We propose to alleviate this problem by ``tricking'' the LRU cache algorithm by updating the timestamp of the data we want to keep in cache (namely entries of matrix $C$). This in effect enables us to have all the benefits of an explicit cache policy while being constrained by the LRU paradigm (realistic policy on a CPU). 
    \end{doublespace}
\end{abstract}

\section{Introduction}
The LRU (Least Recently Used) cache policy follows a FIFO (First In, First Out) policy. This is because, when the cache is full, the first entry that has been inserted into the cache is the one with the lowest ``timestamp'' value (the time at which it was inserted into the cache), and this first entry is the one to be removed. In a similar manner, the LFU (Least Frequently Used) cache policy removes the value in the cache which has been used the least (it has the lowest frequency count). If there is a tie, meaning multiple elements in the cache have been used the same least number of times, the tiebreaker is determined by the ``oldest'' value, or the one with the smallest timestamp value. Although there are strict lower bounds for communication for matrix-matrix multiplication, these assume \textbf{explicit cache control}, meaning one can pick and choose at will what entries of the cache are the least useful and should be removed in any given scenario. In practice, however, it is difficult to implement an explicit cache control, because this would require giving a computer the ability to know precisely which entries of the cache to keep and which to evict at every possible situation. This is why cache policies such as LRU and LFU are standard. The purpose of this paper is to utilize the Cache Simulator to count the communication for algorithms using LRU and LFU (and, for purely theoretical purposes, explicit cache control, for which we are able to simulate using a manipulation of our LRU cache function as will be explained). The importance of minimizing communication between the main memory and cache memory is twofold. Firstly, computing energy is saved, because the computer processor simply has to perform less operations (less exchanges) between data in the cache and main memory. Secondly, the time taken to reach the solution is vastly decreased. For small values of $n$ and $M$, the differences are negligible, but for large $n$ and $M$, which are used in practice, the differences are significant.

\section{Matrix-Matrix Multiplication Defined}
The standard matrix-matrix multiplication algorithm is well known and involves ``row by column'' operations. For example, take 
\begin{center}
A = 
$\begin{bmatrix}
a_{11} & a_{12} \\
a_{21} & a_{22}
\end{bmatrix}
\times
B = 
\begin{bmatrix}
b_{11} & b_{12} \\
b_{21} & b_{22}
\end{bmatrix}$
\end{center}
In order to compute the first entry of the product $C$, we look at the first row of $A$ and multiply each element by the corresponding elements of the first column of $B$. So 
$C_{11} = a_{11} \cdot b_{11} + a_{12} \cdot b_{22}$.
Likewise, we can compute all the elements of $C$:
\begin{center}
C = 
$\begin{bmatrix}
a_{11} \cdot b_{11} + a_{12} \cdot b_{22} & a_{11} \cdot b_{12} + a_{12} \cdot b_{22}\\
a_{21} \cdot b_{11} + a_{22} \cdot b_{22} & a_{21} \cdot b_{12} + a_{22} \cdot b_{22}
\end{bmatrix}$
\end{center}
In general, we are looking at $n \times n$ matrices $A$ and $B$.  
\begin{center}
$
A = \begin{bmatrix}
a_{11} & a_{12} & \cdots & a_{1n} \\
a_{21} & a_{22} & \cdots & a_{2n} \\
\vdots & \vdots & \ddots & \vdots \\
a_{n1} & a_{n2} & \cdots & a_{nn}
\end{bmatrix}
\times
B = \begin{bmatrix}
b_{11} & b_{12} & \cdots & b_{1n} \\
b_{21} & b_{22} & \cdots & b_{2n} \\
\vdots & \vdots & \ddots & \vdots \\
b_{n1} & b_{n2} & \cdots & b_{nn}
\end{bmatrix}
$
\end{center}
So
$C_{11} = a_{11} \cdot b_{11} + a_{12} \cdot b_{21} + \cdots + a_{1n} \cdot b_{n1}$, or $\sum_{i=1}^{n} a_{1i} \cdot b_{i1}$.
The entire representation of $C$ is as follows, and is described by the following algorithm, where we are moving horizontally (across a row) of $A$ and vertically (down a column) of $B$:
\begin{verbatim}
for (int i = 0; i < n; i++) 
  for (int j = 0; j < n; j++) 
    for (int k = 0; k < n; k++) 
      C[i][j] = C[i][j] + A[i][k] * B[k][j];
\end{verbatim}
\begin{center}
$
C = \begin{bmatrix}
\sum_{i=1}^{n} a_{1i} \cdot b_{i1} & \sum_{i=1}^{n} a_{1i} \cdot b_{i2} & \cdots & \sum_{i=1}^{n} a_{1i} \cdot b_{in} \\
\sum_{i=1}^{n} a_{2i} \cdot b_{i1} & \sum_{i=1}^{n} a_{2i} \cdot b_{i2} & \cdots & \sum_{i=1}^{n} a_{2i} \cdot b_{in} \\
\vdots & \vdots & \ddots & \vdots \\
\sum_{i=1}^{n} a_{ni} \cdot b_{i1} & \sum_{i=1}^{n} a_{ni} \cdot b_{i2} & \cdots & \sum_{i=1}^{n} a_{ni} \cdot b_{in}
\end{bmatrix}
$
\end{center}

\section{I/O data movement}

\subsection{Model: LRU (Least Recently Used) Cache}
I/O is represented by input/output, and this numerical quantity is communication. Below, we will discuss I/O as the analogous form reads/writes for a $4 \times 4$ matrix-matrix multiplication case. Reads represents the number of new data values the cache reads in from the main memory, while writes represents the number of data values the cache must write back to the main memory. The cache is a small, 1-D set of data values but the main memory is essentially infinite in comparison. First, we briefly illustrate the LRU Cache Policy.
\newline
\newline
Below a full cache of size 4 is depicted:

\begin{center}
\begin{tabular}{|c|c|c|c|c|}
\hline
\text{Cache entry} & {$a$} & \text{$b$} & \text{$c$} & \text{$d$} \\
\hline
\text{timestamp} & {0} & \text{1} & \text{2} & \text{3} \\
\hline
\end{tabular}
\end{center}

Now assume we want to bring a new element $e$ into our cache. With an LRU cache policy, we will evict the entry with the lowest timestamp. So after $e$ is called, we will increment our read counter by 1, and our cache entry and timestamp arrays will look like: 
\begin{center}
\begin{tabular}{|c|c|c|c|c|}
\hline
\text{Cache entry} & {$e$} & \text{$b$} & \text{$c$} & \text{$d$} \\
\hline
\text{timestamp} & {4} & \text{1} & \text{2} & \text{3} \\
\hline
\end{tabular}
\end{center}
\subsection{Example blocking of 4x4 matrix-matrix multiplication }

We must first understand why blocked matrix-matrix multiplication works and is superior to the standard algorithm for matrix-matrix multiplication when thinking about an optimal cache. It is discussed in a paper from Dongarra, Pineau, Robert, Shi, and Vivien~\cite{dongarra-2008}. Additionally, we discuss the general superiority in communication of Explicit Cache Control compared to the LRU Cache policy.
\newline
\newline
Let us assume we have three $4 \times 4$ matrices, $A$, $B$, and $C$.
\begin{center}
$A = \begin{bmatrix}
a_{00} & a_{01} & a_{02} & a_{03} \\
a_{10} & a_{11} & a_{12} & a_{13} \\
a_{20} & a_{21} & a_{22} & a_{23} \\
a_{30} & a_{31} & a_{32} & a_{33}
\end{bmatrix}$
$\times$
$B = \begin{bmatrix}
b_{00} & b_{01} & b_{02} & b_{03} \\
b_{10} & b_{11} & b_{12} & b_{13} \\
b_{20} & b_{21} & b_{22} & b_{23} \\
b_{30} & b_{31} & b_{32} & b_{33}
\end{bmatrix}$
=
$C = \begin{bmatrix}
c_{00} & c_{01} & c_{02} & c_{03} \\
c_{10} & c_{11} & c_{12} & c_{13} \\
c_{20} & c_{21} & c_{22} & c_{23} \\
c_{30} & c_{31} & c_{32} & c_{33}
\end{bmatrix}$
\end{center}

Now we will ``block'' $A$, $B$, and $C$ by cutting them each into four $2 \times 2$ blocks: 
\[
A = \begin{bmatrix}
a_{00} & a_{01} & \vline & a_{02} & a_{03} \\
a_{10} & a_{11} & \vline & a_{12} & a_{13} \\
\hline
a_{20} & a_{21} & \vline & a_{22} & a_{23} \\
a_{30} & a_{31} & \vline & a_{32} & a_{33}
\end{bmatrix}
\]
\[
B = \begin{bmatrix}
b_{00} & b_{01} & \vline & b_{02} & b_{03} \\
b_{10} & b_{11} & \vline & b_{12} & b_{13} \\
\hline
b_{20} & b_{21} & \vline & b_{22} & b_{23} \\
b_{30} & b_{31} & \vline & b_{32} & b_{33}
\end{bmatrix}
\]
\[
C = \begin{bmatrix}
c_{00} & c_{01} & \vline & c_{02} & c_{03} \\
c_{10} & c_{11} & \vline & c_{12} & c_{13} \\
\hline
c_{20} & c_{21} & \vline & c_{22} & c_{23} \\
c_{30} & c_{31} & \vline & c_{32} & c_{33}
\end{bmatrix}
\]
To demonstrate the importance of blocked matrix-matrix multiplication in conjunction with a limited cache, let us assume we want to compute the 4 entries in the upper left quadrant of $C$: $c_{00}, c_{01}, c_{10}, c_{11}$. Note that this would require using all data from the first two rows of $A$ and the first two columns of $B$. Considering the entries in the upper left quadrant of $A$ and $B$, observe that $a_{00}$ is needed to perform the following two multiplications: $a_{00} \cdot b_{00}$
and $a_{00} \cdot b_{01}$, meaning that $a_{00}$ can be reused once. Similarly, all other values in the upper left quadrant of $A$ can be reused. For $c_{00} = \sum_{i=0}^{3} a_{0i} \cdot b_{i0}$, we can utilize the left quadrant blocks of $A$ and $B$ to compute half the required terms ($a_{00} \cdot b_{00}$ and $a_{00} \cdot b_{01}$ are the 2 multiplications we can perform out of a total of 4 needed for $c_{00}$). Our algorithm will store $a_{00} \cdot b_{00} + a_{00} \cdot b_{01}$ in $c_{00}$ and then we can write back $c_{00}$ to the main memory before retrieving it later to finish calculating the sum of the final two terms: $a_{00} \cdot b_{02} + a_{00} \cdot b_{03}$. So, for a blocking of size $2 \times 2$ as opposed to $1 \times 1$ (which is just the standard algorithm), we have halved the number of reads of data values from the main memory. For Explicit Cache Control, it is easy to see that we aim to keep $C$ entries of the block we are working on in cache during the computation process. For LRU, we cannot control the fact that eventually a $C$ entry will have the lowest timestamp and be kicked out of the cache. Following this, our algorithm will have to perform more reads and writes to recall $C$ and finish the computation. 
\newline
\newline
The cache-efficient way of performing matrix-matrix multiplication (blocking) is illustrated by the following six-looped advanced algorithm, where all three $i$, $j$, and $k$ values increment by different values ($b_i, b_j,$ and $b_k$, respectively).
\begin{verbatim}
for (i = 0; i < n; i += bi) 
    for (j = 0; j < n; j += bj) 
        for (k = 0; k < n; k += bk) 
            for (ib = i; (ib < i + bi) && (ib < n); ib++) 
                for (jb = j; (jb < j + bj) && (jb < n); jb++) 
                    for (kb = k; (kb < k + bk) && (kb < n); kb++) 
                        C[ib][jb] += A[ib][kb] * B[kb][jb];
\end{verbatim}
\subsection{The $\mathcal{O}(M)$ Cache Simulator}
Below is the full code representing our $\mathcal{O}(M)$ cache simulator. The table given in the example explains how it works. Basically, if the element we are looking for is in cache, then simply update its timestamp. If the element is not in cache, find the current oldest variable in the cache and replace it with this element (this element will become the youngest, as it is the newest one used).
\begin{verbatim}
for (i = 0; i < n; i += bi) 
    for (j = 0; j < n; j += bj) 
        for (k = 0; k < n; k += bk) 
            for (ib = i; (ib < i + bi) && (ib < n); ib++) 
                for (jb = j; (jb < j + bj) && (jb < n); jb++) 
                    for (kb = k; (kb < k + bk) && (kb < n); kb++) 
                        C[ib][jb] += A[ib][kb] * B[kb][jb];
cache_functionLRU(M, &time, 0, Aid[ib][kb], isincache, timestamps, changed, &read, &write);
cache_functionLRU(M, &time, 0, Bid[kb][jb], isincache, timestamps, changed, &read, &write);
cache_functionLRU(M, &time, 1, Cid[ib][jb], isincache, timestamps, changed, &read, &write);
\end{verbatim}
\hrule
\begin{verbatim}
int cache_functionLRU (int M, int *time, int ischanged, int id, 
						int *isincache, int *timestamps, int *changed,
						int *read, int *write) {
				(*time)++;
				int itisincache = 0;
				for (int p = 0; p < M; p++) {
					if (id == isincache[p]) {
						timestamps[p] = (*time);
						itisincache = 1;
						changed[p] = ischanged;
					} 
				}
				if (itisincache == 0) {
					int oldest_time = (*time);
					int oldest_loc = M;
					for (int p = 0; p < M; p++) {
						if (timestamps[p] < oldest_time) {
							oldest_time = timestamps[p];
							oldest_loc = p;
						}
					}
					if (changed[oldest_loc] == 1) (*write)++;
					(*read)++;
					isincache[oldest_loc] = id;
					changed[oldest_loc] = ischanged;
					timestamps[oldest_loc] = (*time);
				}
				return 0;
			}
for (int p = 0; p < M; p++) if (write_or_no[p] == 1) write++;	
\end{verbatim}
\section{Example}
This example table with $n = 4$ and $M = 12$ demonstrates the LRU cache policy in action by displaying elements in cache and their timestamp values. Blue correlates with data reuse, red with a data read, and purple with a data read and write.
$$
\begin{array}{|c|c|c|c|c|c|c|c|c|c|c|c|c|c|} \hline 
initial & c_{00} = c_{00} + a_{00}b_{00} & 
\varnothing & \varnothing & \varnothing &
\varnothing & \varnothing & \varnothing &
\varnothing & \varnothing & \varnothing &
\varnothing & \varnothing & \varnothing 
\\\hline 
&
r = 0, w = 0
&
0 & 1 & 2 & 
3 & 4 & 5 &
6 & 7 & 8 &
9 & 10 & 11 \\\hline 
n_{000} & c_{00} = c_{00} + a_{00}b_{00} & 
\color{red} a_{00} & \color{red}b_{00} & \color{red}c_{00} &
\varnothing & \varnothing & \varnothing &
\varnothing & \varnothing & \varnothing &
\varnothing & \varnothing & \varnothing  \\\hline 
&
r = 3, w = 0
&
\color{red}12 & \color{red}13 & \color{red}14 & 
3 & 4 & 5 &
6 & 7 & 8 &
9 & 10 & 11 \\\hline 

n_{001} & c_{00} = c_{00} + a_{01}b_{10} & 
a_{00} & b_{00} & \color{blue}c_{00} & 
\color{red}a_{01} & \color{red}b_{10} & \varnothing & 
\varnothing & \varnothing & \varnothing & 
\varnothing & \varnothing & \varnothing  
\\\hline 
&
r = 5, w = 0
&
12 & 13 & \color{blue}17 & \color{red}15 & \color{red}16 & 5 & 6 & 7 & 8 & 9 & 10 & 11 \\\hline 

n_{002} & c_{00} = c_{00} + a_{02}b_{20} & 
a_{00} & b_{00} & \color{blue}c_{00} & 
a_{01} & b_{10} & \color{red}a_{02} & 
\color{red}b_{20} & \varnothing & \varnothing &
\varnothing & \varnothing & \varnothing 

\\\hline 
&
r = 7, w = 0
&
12 & 13 & \color{blue}20 & 15 & 16 & \color{red}18 & \color{red}19 & 7 & 8 & 9 & 10 & 11 \\\hline
n_{003} & c_{00} = c_{00} + a_{03}b_{30} & 
a_{00} & b_{00} & \color{blue}c_{00} & 
a_{01} & b_{10} & a_{02} & 
b_{20} & \color{red}a_{03} & \color{red}b_{30} &
\varnothing & \varnothing & \varnothing 
\\\hline 
&
r = 9, w = 0
&
12 & 13 & \color{blue}23 & 15 & 16 & 18 & 19 & \color{red}21 & \color{red}22 & 9 & 10 & 11 \\\hline  
n_{010} & c_{01} = c_{01} + a_{00}b_{01} & 
\color{blue}a_{00} & b_{00} & c_{00} & 
a_{01} & b_{10} & a_{02} & 
b_{20} & a_{03} & b_{30} & 
\color{red}b_{01} & \color{red}c_{01} & \varnothing 
\\\hline 
&
r = 11, w = 0
&
\color{blue}24 & 13 & 23 & 15 & 16 & 18 & 19 & 21 & 22 & \color{red}25 & \color{red}26 & 11 \\\hline 
n_{011} & c_{01} = c_{01} + a_{01}b_{11} & 
a_{00} & b_{00} & c_{00} & 
\color{blue}a_{01} & b_{10} & a_{02} & 
b_{20} & a_{03} & b_{30} & 
b_{01} & \color{blue}c_{01} & \color{red}b_{11}
\\\hline 
&
r = 12, w = 0
&
24 & 13 & 23 & \color{blue}27 & 16 & 18 & 19 & 21 & 22 & 25 & \color{blue}29 & \color{red}28 \\\hline 
n_{012} & c_{01} = c_{01} + a_{02}b_{21} & 
a_{00} & \color{red}b_{21} & c_{00} & 
a_{01} & b_{10} & \color{blue}a_{02} & 
b_{20} & a_{03} & b_{30} & 
b_{01} & \color{blue}c_{01} & b_{11} 
\\\hline 
&
r = 13, w = 0
&
24 & \color{red}31 & 23 & 27 & 16 & \color{blue}30 & 19 & 21 & 22 & 25 & \color{blue}32 & 28 \\\hline  
n_{013} & c_{01} = c_{01} + a_{03}b_{31} & 
a_{00} & b_{21} & c_{00} & 
a_{01} & \color{red}b_{31} & a_{02} & 
b_{20} & \color{blue}a_{03} & b_{30} & 
b_{01} & \color{blue}c_{01} & b_{11}
\\\hline 
&
r = 14, w = 0
&
24 & 31 & 23 & 27 & \color{red}34 & 30 & 19 & \color{blue}33 & 22 & 25 & \color{blue}35 & 28 \\\hline 
n_{020} & c_{02} = c_{02} + a_{00}b_{02} & 
\color{blue}a_{00} & b_{21} & c_{00} & 
a_{01} & b_{31} & a_{02} & 
\color{red}b_{02} & a_{03} & \color{red}c_{02} & 
b_{01} & c_{01} & b_{11} 
\\\hline 
&
r = 16, w = 0
&
\color{blue}36 & 31 & 23 & 27 & 34 & 30 & \color{red}37 & 33 & \color{red}38 & 25 & 35 & 28 \\\hline 
n_{021} & c_{02} = c_{02} + a_{01}b_{12} & 
a_{00} & b_{21} & \color{purple}b_{12} & 
\color{blue}a_{01} & b_{31} & a_{02} & 
b_{02} & a_{03} & \color{blue}c_{02} & 
b_{01} & c_{01} & b_{11} 
\\\hline 
&
r = 17, w = 1
&
36 & 31 & \color{purple}40 & \color{blue}39 & 34 & 30 & 37 & 33 & \color{blue}41 & 25 & 35 & 28 \\\hline 
n_{022} & c_{02} = c_{02} + a_{01}b_{12} & 
a_{00} & b_{21} & b_{12} & 
a_{01} & b_{31} & \color{blue}a_{02} & 
b_{02} & a_{03} & \color{blue}c_{02} & 
\color{red}b_{22} & c_{01} & b_{11} 
\\\hline 
&
r = 18, w = 1
&
36 & 31 & 40 & 39 & 34 & \color{blue}42 & 37 & 33 & \color{blue}44 & \color{red}43 & 35 & 28 \\\hline 
n_{023} & c_{02} = c_{02} + a_{03}b_{32} & 
a_{00} & b_{21} & b_{12} & 
a_{01} & b_{31} & a_{02} & 
b_{02} & \color{blue}a_{03} & \color{blue}c_{02} & 
b_{22} & c_{01} & \color{red}b_{32}
\\\hline 
&
r = 19, w = 1
&
36 & 31 & 40 & 39 & 34 & 42 & 37 & \color{blue}45 & \color{blue}47 & 43 & 35 & \color{red}46 \\\hline

n_{030} & c_{03} = c_{03} + a_{00}b_{03} & 
\color{blue}a_{00} & \color{red}b_{03} & b_{12} & 
a_{01} & \color{red}c_{03} & a_{02} & 
b_{02} & a_{03} & c_{02} & 
b_{22} & c_{01} & b_{32} 
\\\hline 
&
r = 21, w = 1
&
\color{blue}48 & \color{red}49 & 40 & 39 & \color{red}50 & 42 & 37 & 45 & 47 & 43 & 35 & 46 \\\hline

n_{031} & c_{03} = c_{03} + a_{01}b_{13} & 
a_{00} & b_{03} & b_{12} & 
\color{blue}a_{01} & \color{blue}c_{03} & a_{02} & 
b_{02} & a_{03} & c_{02} & 
b_{22} & \color{purple}b_{13} & b_{32} 
\\\hline 
&
r = 22, w = 2
&
48 & 49 & 40 & \color{blue}51 & \color{blue}53 & 42 & 37 & 45 & 47 & 43 & \color{purple}52 & 46\\\hline

n_{032} & c_{03} = c_{03} + a_{01}b_{13} & 
a_{00} & b_{03} & b_{12} & 
a_{01} & \color{blue}c_{03} & \color{blue}a_{02} & 
\color{red}b_{23} & a_{03} & c_{02} & 
b_{22} & b_{13} & b_{32}
\\\hline 
&
r = 23, w = 2
&
48 & 49 & 40 & 51 & \color{blue}56 & \color{blue}54 & \color{red}55 & 45 & 47 & 43 & 52 & 46\\\hline

n_{033} & c_{03} = c_{03} + a_{01}b_{13} & 
a_{00} & b_{03} & \color{red}b_{33} & 
a_{01} & \color{blue}c_{03} & a_{02} & 
b_{23} & \color{blue}a_{03} & c_{02} & 
b_{22} & b_{13} & b_{32} 
\\\hline 
&
r = 24, w = 2 
&
48 & 49 & \color{red}58 & 51 & \color{blue}59 & 54 & 55 & \color{blue}57 & 47 & 43 & 52 & 46\\\hline

n_{100} & c_{00} = c_{10} + a_{10}b_{00} & 
a_{00} & b_{03} & b_{33} & 
a_{01} & c_{03} & a_{02} & 
b_{23} & a_{03} & \color{purple}c_{10} & 
\color{red}a_{10} & b_{13} & \color{red}b_{00} 
\\\hline 
&
r = 27, w = 3
&
48 & 49 & 58 & 51 & 59 & 54 & 55 & 57 & \color{purple}62 & \color{red}60 & 52 & \color{red}61\\\hline

n_{101} & c_{10} = c_{10} + a_{11}b_{10} & 
\color{red}a_{11} & \color{red}b_{10} & b_{33} & 
a_{01} & c_{03} & a_{02} & 
b_{23} & a_{03} & \color{blue}c_{10} & 
a_{10} & b_{13} & b_{00} 
\\\hline 
&
r = 29, w = 3
&
\color{red}63 & \color{red}64 & 58 & 51 & 59 & 54 & 55 & 57 & \color{blue}65 & 60 & 52 & 61\\\hline

n_{102} & c_{10} = c_{10} + a_{12}b_{20} & 
a_{11} & b_{10} & b_{33} & 
\color{red}a_{12} & c_{03} & a_{02} & 
b_{23} & a_{03} & \color{blue}c_{10} & 
a_{10} & \color{red}b_{20} & b_{00} 
\\\hline 
&
r = 31, w = 3 
&
63 & 64 & 58 & \color{red}66 & 59 & 54 & 55 & 57 & \color{blue}68 & 60 & \color{red}67 & 61\\\hline

n_{103} & c_{10} = c_{10} + a_{13}b_{30} & 
a_{11} & b_{10} & b_{33} & 
a_{12} & c_{03} & \color{red}a_{13} & 
\color{red}b_{30} & a_{03} & \color{blue}c_{10} & 
a_{10} & b_{20} & b_{00} 
\\\hline 
&
r = 33, w = 3 
&
63 & 64 & 58 & 66 & 59 & \color{red}69 & \color{red}70 & 57 & \color{blue}71 & 60 & 67 & 61\\\hline

n_{110} & c_{11} = c_{11} + a_{10}b_{01} & 
\color{blue}a_{11} & b_{10} & \color{red}c_{11} & 
a_{12} & \color{purple}b_{11} & a_{13} & 
b_{30} & b_{01} & c_{10} & 
a_{10} & b_{20} & b_{00} 
\\\hline 
&
r = 35, w = 4 
&
\color{blue}75 & 64 & \color{red}77 & 66 & \color{purple}76 & 69 & 70 & 73 & 71 & 72 & 67 & 61\\\hline

\end{array}\\
$$

\subsection{Asymptotic behavior and Optimal Cache Algorithm}
We will use both graphical and theoretical means to confirm lower bounds of communication assuming explicit cache control. First, it is useful to cite some results about communication lower bounds for matrix-matrix multiplication. Our goal is to have the amount of communication be as small as possible. The next two results give lower bounds on the minimum amount of communication for any matrix-matrix multiplication algorithm. In other words, no matrix-matrix multiplication algorithm can perform better than these bounds. We will see that, for a correct $b$, the blocked algorithm that we call $b$-$b$-1 gets close to the lower bound and therefore is optimal.


\begin{theorem}
Given a cache of size $M$ and matrix-matrix multiplication of $A \times B = C$, where $A$ is $m \times k$, $B$ is $k \times n$, and $C$ is $m \times n$, Hong and Kung~\cite{hong.81.stoc} proved that the minimum lower bound of communication is asymptotically on the order of $\frac{2mnk}{\sqrt{M}}$.
\end{theorem}

Another proof for the same result was given in Irony, Toledo, and Tiskin~\cite{toledo.jpdc}.
Olivry et al.~\cite{olpsr:pldi20:20} give a nonasymptotic results.

\begin{theorem}
Olivry et al.~\cite{olpsr:pldi20:20} states that, for three $n\times n$ $A$, $B$, $C$ matrices and 
a cache of size $M$, then
any matrix-matrix multiplication algorithm
needs to perform a volume of communication (IO) 
according to the following restriction
$$\textmd{IO} \geq \frac{2n^3}{\sqrt{M}} - \frac{2n^2}{\sqrt{M}} + 5n - M - 2$$.
\end{theorem}

\begin{theorem}
Assume a six-loop blocking with $b_i = b_j = b_k = b$,
the largest $b$ such that three tiles (one of $A$, one of $B$ and one of $C$)
fit in cache is 
$$ b = \lfloor \frac{\sqrt{M}}{\sqrt{3}}\rfloor . $$
\end{theorem}
{\bf Proof:} 
For three $b\times b$ tiles to fit in cache of size $M$, we need
$$ 3 b^2 \leq M. $$
This gives 
$$ b \leq \frac{\sqrt{M}}{\sqrt{3}}. $$
So we want
$$ b = \lfloor \frac{\sqrt{M}}{\sqrt{3}}\rfloor . $$

\begin{theorem}
Assume a six-loop blocking with $b_i = b_j = b$ and $b_k=1$,
the largest $b$ such that three tiles (one of $A$, one of $B$ and one of $C$)
fit in cache is 
$$ b = \lfloor \sqrt{M+1}\rfloor -1 . $$
\end{theorem}
{\bf Proof:} 
For one $b\times b$ tile of $C$,
and one $b\times 1$ tile of $A$,
and one $1\times b$ tile of $B$ to fit in
cache of size $M$, we need
$$ b^2 + 2b \leq M. $$
This gives 
$$b = \lfloor \sqrt{M+1}\rfloor -1$$

\begin{theorem}
Assume a six-loop blocking with $b_i = b_j = b$ and $b_k=\alpha$, where $\alpha$ is fixed,
the largest $b$ such that three tiles (one of $A$, one of $B$ and one of $C$)
fit in cache is 
$$ b = -\alpha + \sqrt{\alpha^2 + M} $$
\end{theorem}
{\bf Proof:} 
For one $b\times b$ tile of $C$,
and one $b\times k$ tile of $A$,
and one $k\times b$ tile of $B$ to fit in
cache of size $M$, we need
$$ b^2 + 2b\alpha \leq M. $$
This gives 
$$b \leq -\alpha + \sqrt{\alpha^2 + M}$$

\begin{theorem}
Assume a six-loop blocking with $b_i = b_j = b_k = b$,
then the number of I/O (assuming explicit cache control) is 
$$IO = 2\sqrt{3}\frac{n^3}{\sqrt{M}} + n^2 $$
\end{theorem}

{\bf Proof:} 
Using our algorithm for matrix-matrix multiplication:
\begin{Verbatim}[commandchars=\\\{\}]
for (i = 0; i < n; i += bi) 
    for (j = 0; j < n; j += bj) 
\textcolor{blue}{//      load C(i:i+bi-1,j:j+jb-1)}
        for (k = 0; k < n; k += bk) 
\textcolor{blue}{//          load A(i:i+bi-1,k:k+kb-1)}
\textcolor{blue}{//          load B(k:k+bk-1,j:j+jb-1)}
            for (ib = i; (ib < i + bi) && (ib < n); ib++) 
                for (jb = j; (jb < j + bj) && (jb < n); jb++) 
                    for (kb = k; (kb < k + bk) && (kb < n); kb++) 
                        C[ib][jb] += A[ib][kb] * B[kb][jb];
\end{Verbatim}

Assuming that $b \mid n$ (blocks of size $b$ tile the $n \times n$ matrix):
\begin{eqnarray}
\nonumber & = &    
 \left(\sum_{i=1}^{n/b}\sum_{j=1}^{n/b} b^2 \right)
+ 2 \left(\sum_{i=1}^{n/b}\sum_{j=1}^{n/b}\sum_{k=1}^{n/b} b^2 \right)\\
\nonumber & = &    
\left(\frac{n}{b}\right)^2 b^2 + 
2\left(\frac{n}{b}\right)^3 b^2 \\
\nonumber & = &    
n^2 + \frac{2n^3}{b}\\
\nonumber & = &    
n^2 + \frac{2n^3\sqrt{3}}{\sqrt{M}}\\
\nonumber & = &    
2\sqrt{3}\frac{n^3}{\sqrt{M}} + n^2
\end{eqnarray}

\begin{theorem}
Assume a six-loop blocking with $b_i = b_j = b $ and $ b_k = 1$,
then the number of I/O (assuming explicit cache control) is 
$$IO = 2\frac{n^3}{\sqrt{M}} + n^2 $$
\end{theorem}
{\bf Proof:} 
Once again, assuming that $b \mid n$,
\begin{eqnarray}
\nonumber & = &    
 \left(\sum_{i=1}^{n/b}\sum_{j=1}^{n/b} b^2 \right)
+ 2 \left(\sum_{i=1}^{n/b}\sum_{j=1}^{n/b}\sum_{k=1}^{n/b} b^2 \right)\\
\nonumber & = &    
\left(\frac{n}{b}\right)^2 b^2 + 
2\left(\frac{n}{b}\right)^3 b^2 \\
\nonumber & = &    
n^2 + \frac{2n^3}{b}\\
\nonumber & = &    
n^2 + \frac{2n^3}{\lfloor \sqrt{M+1}\rfloor -1}\\
\nonumber & \approx &    
\frac{2n^3}{\sqrt{M}} + n^2
\end{eqnarray}
For large n, we arrive at
$\frac{2n^3}{\sqrt{M}}$
as confirmed in a paper from Smith, Lowery, Langou, and Van De Geijn~\cite{sllv:arxiv:19}.
\begin{theorem}
Assume a six-loop blocking with $b_i = b_j = b $ and $ b_k = \alpha$,
then the number of I/O (assuming explicit cache control) is 
$$IO = \frac{2n^3}{-\alpha + \sqrt{\alpha^2 + M}} + n^2$$
\end{theorem}
{\bf Proof:} 
Once again, assuming that $b \mid n$,
\begin{eqnarray}
\nonumber & = &    
 \left(\sum_{i=1}^{n/b}\sum_{j=1}^{n/b} b^2 \right)
+ 2 \left(\sum_{i=1}^{n/b}\sum_{j=1}^{n/b}\sum_{k=1}^{n/b} b^2 \right)\\
\nonumber & = &    
\left(\frac{n}{b}\right)^2 b^2 + 
2\left(\frac{n}{b}\right)^3 b^2 \\
\nonumber & = &    
n^2 + \frac{2n^3}{b}\\
\nonumber & = &    
n^2 + \frac{2n^3}{-\alpha + \sqrt{\alpha^2 + M}}\\
\nonumber & = &    
\frac{2n^3}{-\alpha + \sqrt{\alpha^2 + M}} + n^2
\end{eqnarray}


\begin{theorem}
Assuming explicit cache control, a matrix of size $n \times n$, and a cache of size $M$, 
the configuration (bi, bj, bk) for six looped matrix-matrix multiplication that will perform the least amount of I/O is the arrangement ($b$, $b$, 1).
\end{theorem}

{\bf Proof:} 
 Any combination of $b_i = b_j \neq b$ is sub-optimal. This is shown graphically below. Additionally, we have shown that, for $b_k = \alpha$, the number of I/O is 
$$IO = \frac{2n^3}{-\alpha + \sqrt{\alpha^2 + M}} + n^2$$
\newline
\newline
Therefore, we want to find $\alpha$ so that our I/O expression is minimized. Clearly, if $\alpha = 1$, the expression $\alpha + \sqrt{\alpha^2 + M}$ is closest to $\sqrt{M}$. As $\alpha$ is increased, $\alpha + \sqrt{\alpha^2 + M}$ approaches 0, and I/O will increase. So the best choice of $\alpha$ is 1, and our I/O is $\approx 
\frac{2n^3}{\sqrt{M}} + n^2$. 

\section{A Discussion of Other Cache Policies}
Although we have extensively covered the LRU cache policy, we briefly explain how we were able to simulate our theoretical results using the explicit cache policy. Additionally, we touch on the LFU cache policy. 
\subsection{Explicit Cache Control}
In order to simulate explicit cache control, one can add the following nested for-loops just outside the inner most for-loop (referring back to the $\mathcal{O}(M)$ Cache Simulator):
\begin{Verbatim}
for(int ib1 = i; (ib1 < i + bi)&&(ib1 < n); ib1++)
    for(int jb1 = j; (jb1 < j + bj)&&(jb1 < n); jb1++) 
	LRU(M, &time, 1, Cid[ib1][jb1], isincache, timestamps, changed, &read, &write);
\end{Verbatim}
This simulates explicit cache control because we are simply updating the ``timestamp" values of each $C$ element in our current block. This ensures that the $C$ is not discarded from the cache according to the LRU cache policy, resulting in a drastic drop in reads and writes.
\subsection{LFU Cache policy}
Below a full cache of size 4 is depicted:
\begin{center}
\begin{tabular}{|c|c|c|c|c|}
\hline
\text{Cache entry} & {$a$} & \text{$b$} & \text{$c$} & \text{$d$} \\
\hline
\text{frequency stamp} & {1} & \text{2} & \text{3} & \text{4} \\
\hline
\end{tabular}
\end{center}
Now assume we want to bring a new element $e$ into our cache. With an LRU cache policy, we will evict the entry with the lowest ``frequency stamp" (it has been used the least). So after $e$ is called, our cache entry and frequency stamp arrays will look like: 
\begin{center}
\begin{tabular}{|c|c|c|c|c|}
\hline
\text{Cache entry} & {$e$} & \text{$b$} & \text{$c$} & \text{$d$} \\
\hline
\text{frequency stamp} & {1} & \text{1} & \text{2} & \text{3} \\
\hline
\end{tabular}
\end{center}
A fixed LFU Cache is not ideal for matrix-matrix multiplication, because $C$ values will constantly be removed from the cache due to their frequency being 1 (and once they are removed, their frequency will go back to 0). We note the LFU data curve in figure 2, and omit its code.
\begin{figure}[H]
  \centering
  \includegraphics[width=\textwidth]{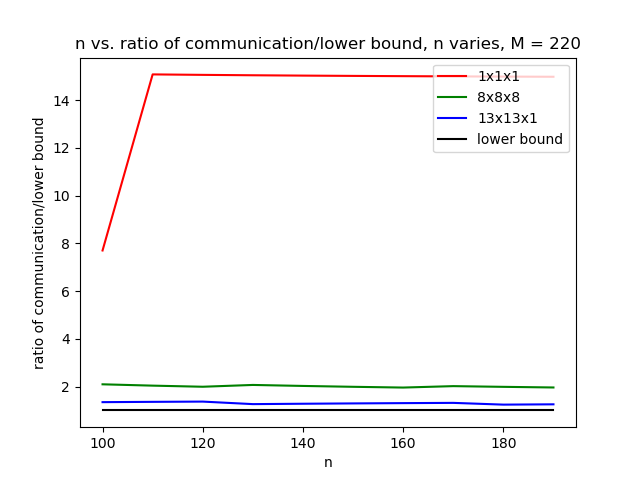}
  \caption{}
  \label{fig:varying values of n}
\end{figure}

\begin{figure}[H]
  \centering
  \includegraphics[width=\textwidth]{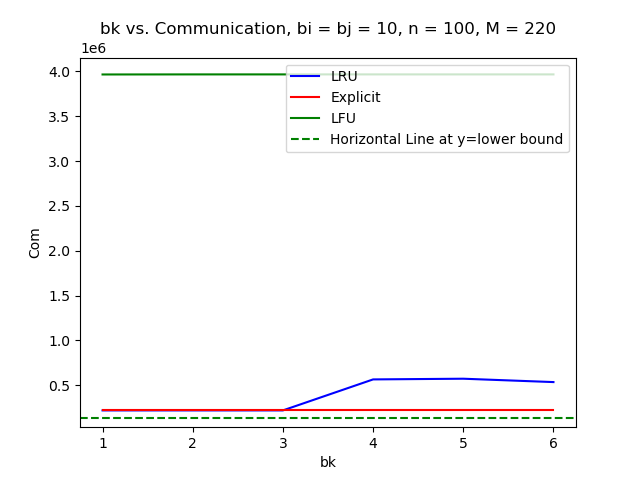}
  \caption{}
  \begin{flushleft}
   \textbf{Figure 1}: Testing communication for the three different matrix-matrix multiplication algorithms with varying values of $n$. All curves have been standardized against the lower bound, which is why the lower bound curve is 1. For our constant $M = 220$, we use our theoretical results to determine optimal $b$ for blocking. We use ($\lfloor \sqrt{221} - 1 \rfloor = 13$ for bxbx1 blocking.
 $\lfloor \sqrt{220/3} \rfloor = 8$ We use for bxbxb blocking. Note how 
 the blue curve approaches 1 (optimal),
 the green curve approaches $\sqrt{3}$ and the red curve approaches $\sqrt{M} = \sqrt{220} = 14.8$. If we increase $M$, the red curve becomes arbitrarily bad. This demonstrates the importance of blocking.
 \newline
 \newline
 
  \textbf{Figure 2}: Testing communication for varying values of $b_k$, the amount the variable k is updated each loop. Note that we use $b_k$ values up to 6 because in that case our blocks of data fit in cache: $b_i$$b_k$ + $b_i$$b_j$ + $b_j$$b_k$ $\leq$ 220. Explicit cache communication values are not ideal, because if they were $b_i = b_j = 13$. Nevertheless, it seems as though small $b_k$ values are best for reducing communication, and explicit cache control is superior to the LRU policy for $b_k > 3$. We chose $b_i = b_j = 10$ to allow $b_k$ to range and our data fit in cache. The LFU Cache Policy seems to show high communication values 
  \label{fig:varying values of bk}
  \end{flushleft}
\end{figure}

\begin{figure}[H]
  \centering
  \includegraphics[width=\textwidth]{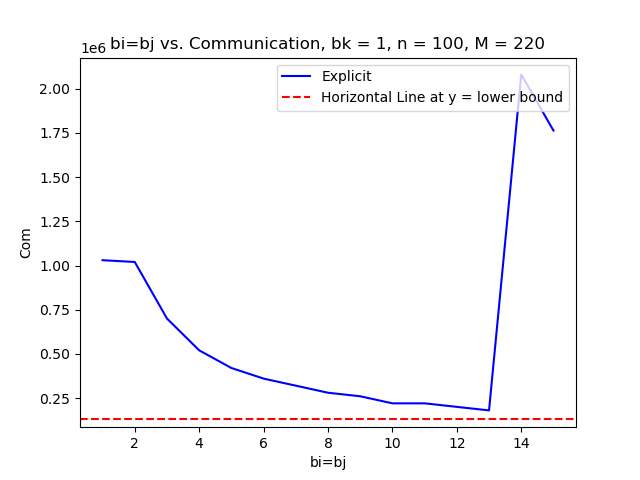}
  \caption{Testing communication for varying values of $b_i=b_j$, the amount the variables $i$ and $j$ are updated each loop. Observe that the value of $b$ for which $b_i = b_j$ has the least communication is 13. This makes sense, because, according to our formula that $b_i = b_j = $ $\lfloor \sqrt{M+1}\rfloor - 1$,  $b = \lfloor \sqrt{220+1}\rfloor -1 = 13$.}
  \label{fig:varying values of bi=bj}
\end{figure}

\section{The $\mathcal{O}(1)$ Cache Simulator}
Now we reveal the $\mathcal{O}(1)$ Cache Simulator, which has the advantage of the algorithmic run time of the LRU cache simulation being independent of $M$. 
\newline
\newline
Here is a detailed explanation of how the $\mathcal{O}(1)$ Cache Simulator works, beginning with an explanation of each function argument and then describing the logic:
\newline
\newline
1. \verb|M| is the size of the cache
\newline
2. \verb|&global_time| is the total number of times the LRU cache function has been called.
\newline
3. \verb|0| or \verb|1|: Note that $A$ and $B$ have values of 0 while $C$ has a value of 1. This value simply indicates that we should not write back values of $A$ or $B$ but should write back values of $C$ the main memory.
\newline
4. \verb|Aid[ib][kb]| or \verb|Bid[kb][jb]| or \verb|Cid[ib][jb]|: this value is the id number given to the entry of the matrix $A$, $B$ or $C$. The way in which we assign id values is demonstrated by:
\begin{Verbatim}
for (i = 0; i < n; i++) {
    for (j = 0; j < n; j++) {
        Aid[i][j] = 0*n*n + n*j + i;
	Bid[i][j] = 1*n*n + n*j + i;
	Cid[i][j] = 2*n*n + n*j + i;
    }
}
\end{Verbatim}
Then, the id values of $A$, $B$, and $C$ are
\begin{center}
$A = \begin{bmatrix}
0 & 1 & 2 & 3 \\
4 & 5 & 6 & 7 \\
8 & 9 & 10 & 11 \\
12 & 13 & 14 & 15
\end{bmatrix}$
$\times$
$B = \begin{bmatrix}
16 & 17 & 18 & 19 \\
20 & 21 & 22 & 23 \\
24 & 25 & 26 & 27 \\
28 & 29 & 30 & 31
\end{bmatrix}$
=
$C = \begin{bmatrix}
32 & 33 & 34 & 35 \\
36 & 37 & 38 & 39 \\
40 & 41 & 42 & 43 \\
44 & 45 & 46 & 47
\end{bmatrix}$
\end{center}

\noindent 5. \verb|id_Array| is an array of size $M$ containing the id values of all the entries in the cache.
\newline
6. \verb|timestamps| is an array of size $M$ containing the \verb|&global_time| values of all the entries in cache.
\newline
7. \verb|write_or_no| is an array of size $M$ containing either the value of 0 or 1 (representing if that entry in cache should be written back or not).
\newline
8. \verb|nextyounger| is an array of size $M$. For a certain entry in the cache, it contains the index of the ``nextyounger" entry to that entry in the cache (the entry with a higher timestamp) value.
\newline
9. \verb|nextolder| is an array of size $M$. For a certain entry in the cache, it contains the index of the ``nextolder" entry to that entry in the cache (the entry with a lower timestamp) value.
\newline
10. \verb|&oldest| is the entry in the cache with the lowest timestamp. This is the entry to be replaced according to the LRU cache policy.
\newline
11. \verb|&youngest| is the entry in the cache with the highest timestamp. The entry that just replaced the previous \verb|&oldest| now assumes the position of the youngest entry in the array.
\newline
12. \verb|&read| is to be incremented in our LRU cache function.
\newline
13. \verb|&write| is to be incremented in our LRU cache function, depending on the \verb|&oldest| entry in the cache was C.
\newline
14. \verb|index_in_cache| is an array of size $3n^2$ which contains the id values of the data entries of $A$, $B$, and $C$ that are in the cache (the other $3n^2 - M$ values of index in cache are set to -1.

\begin{Verbatim}[commandchars=\\\{\}]
for (i = 0; i < n; i += bi) 
    for (j = 0; j < n; j += bj) 
        for (k = 0; k < n; k += bk) 
            for (ib = i; (ib < i + bi) && (ib < n); ib++) 
                for (jb = j; (jb < j + bj) && (jb < n); jb++) 
                    for (kb = k; (kb < k + bk) && (kb < n); kb++) 
                        C[ib][jb] += A[ib][kb] * B[kb][jb];

LRU(M, &global_time, 0, Aid[ib][kb], id_Array, timestamps, write_or_no, nextyounger, 
nextolder, \&oldest, \&youngest, \&read, \&write, index_in_cache);

LRU(M, &global_time, 0, Bid[kb][jb], id_Array, timestamps, write_or_no, nextyounger, 
nextolder, \&oldest, \&youngest, \&read, \&write, index_in_cache);

LRU(M, &global_time, 0, Aid[ib][jb], id_Array, timestamps, write_or_no, nextyounger, 
nextolder, \&oldest, \&youngest, \&read, \&write, index_in_cache);
\end{Verbatim}

\hrule

\begin{Verbatim}
int LRU() {
(*global_time)++;
if (index_in_cache[id] >= 0) {
timestamps[index_in_cache[id]] = (*global_time);
write_or_no[index_in_cache[id]] = write_or_not;
    if (index_in_cache[id] == *oldest) {
	*oldest = nextyounger[index_in_cache[id]];
	nextolder[*oldest] = -1;
	nextyounger[index_in_cache[id]] = -1;
	nextyounger[*youngest] = index_in_cache[id];
	nextolder[index_in_cache[id]] = *youngest;
	*youngest = index_in_cache[id];
	}
    else if (index_in_cache[id] == *youngest) {
	//do nothing
	}
    else {
nextolder[nextyounger[index_in_cache[id]]] = nextolder[index_in_cache[id]];
nextyounger[nextolder[index_in_cache[id]]] = nextyounger[index_in_cache[id]];
nextyounger[index_in_cache[id]] = -1;
nextyounger[*youngest] = index_in_cache[id];
nextolder[index_in_cache[id]] = *youngest;
*youngest = index_in_cache[id];
	}
    }
    if (index_in_cache[id] == -1) {
    if (write_or_no[*oldest] == 1) (*write)++;
    (*read)++;
    if(id_Array[*oldest] >= 0)
    index_in_cache[id_Array[*oldest]] = -1;
    id_Array[*oldest] = id;
    write_or_no[*oldest] = write_or_not;
    timestamps[*oldest] = (*global_time);
    nextolder[*oldest] = *youngest;
    nextyounger[*youngest] = *oldest;
    *youngest = *oldest;
    *oldest = nextyounger[*oldest];
    index_in_cache[id] = *youngest;
    nextolder[*oldest] = -1;
    nextyounger[*youngest] = -1;
    }
    return 0;
}
for (int p = 0; p < M; p++) if (write_or_no[p] == 1) write++;
	io = read + write;	
	elapsed += get_current_time();
\end{Verbatim}
The pseudo-code logic of the $\mathcal{O}(1)$ cache simulator, which uses ``double-chained'' arrays of pointers (nextolder and nextyounger) is as follows:
\newline
1. Increment the global time, which will be assigned to the new youngest value.
\newline
2. Is the index in cache of the data entry of $A$, $B$, or $C$ greater than 0 (or equivalently: is the value we want to use in our current cache?).
\newline
\textbf{If the value is in cache (see 2):}
\newline
3. Update the timestamps and write or no arrays.
\newline
4. This is a special case, namely the value we are searching for is the current oldest. Then tell the current oldest value to point to its current next younger (because its current next younger will be the oldest).
\newline
5. Update the next older of oldest to be -1, since by definition there is not an older element in the cache. By similar logic, since index in cache[id] is the new youngest, there is no younger element so its next younger value is set to -1.
\newline
6. Next younger of the current youngest is set to index in cache[id] which will become the new youngest.
\newline
7. Next older of index in cache[id] is given the current youngest.
8. Finally, the identity of the youngest element changes to index in cache[id].
\newline
9. This is a special case, namely the value we are searching for is the current youngest. This is perfect! We don't have to update pointers to the youngest because it will remain the youngest. In fact, we don't have to do anything.
\newline
10. If the value we are searching for is in our cache, but is not the current youngest or the current oldest, then
\newline
11. The next older of the next younger of index in cache[id] will become the next older of index in cache[id], and the next younger of the next older of index in cache[id] will become the next younger of index in cache[id]. This is because we are shifting the indices of our cache entries by 1. 
\newline
12. Next younger of index in cache[id] is set to -1, as index in cache[id] is set to become the youngest.
\newline
13. Next younger of the current youngest is set to index in cache[id]. Similarly, in the reverse manner, next older of index in cache[id] is set to become the youngest.
\newline
14. Finally, the youngest element points to index in cache[id].
\newline
15. Is the index in cache of the data entry of $A$, $B$, or $C$ greater than 0 (or equivalently: is the value we want to use in our current cache?).
\newline
\textbf{If the value is not in cache (see 2, 15):}
\newline
16. If the value we are searching for is not in cache, then check if the current oldest value in the cache needs to be written back to the main memory (if so, increment write).
\newline
17. Increment read, we have to do this because the oldest value will be kicked out of the cache to make way for this new value.
\newline
18. \verb|if(id_Array[*oldest] >= 0)| is simply a formality needed based on the initialization of the id Array (when all values are -1).
\newline 
19. The next 3 lines update the arrays id Array, write or no, and timestamps. Then next older of the current oldest is set to the youngest (because the current oldest will become the youngest, and next older of this youngest is the current youngest). Similarly, next younger of the current youngest is set to the oldest. 
\newline
20. The pointer of the youngest is set to that of the oldest. 
\newline
21. The pointer to the oldest is set to next younger of the current oldest, because this current oldest will disappear.
\newline
22. Next older of the oldest and next younger of the youngest are set to -1, as expected.
\newline
23. Loop through the remaining entries in the cache, and if there are any $c$ values remaining simply write them back (increment the write). Find the $io$, which is $= read + write$, and update the variable elapsed (which is used for timing) based on the current time. 
\newline
\newline
A demonstration with $n = 2$ and $M = 6$ may be beneficial in order to understand the ``double-chained" array structure. Here is an arrangement of data values of \verb|id_Array|, \verb|timestamps|, \verb|nextyounger|, \verb|nextolder|, and \verb|index_in_cache|, respectively:
\begin{align*}
& \begin{array}{|c|c|c|c|c|c|c|c|c|c|c|c|}
\hline
0 & 4 & 8 & 2 & 5 & 6 \\
\hline
\end{array} \\
& \begin{array}{|c|c|c|c|c|c|c|c|c|c|c|c|}
\hline
12 & 7 & 11 & 9 & 10 & 13 \\
\hline
\end{array} \\
& \begin{array}{|c|c|c|c|c|c|c|c|c|c|c|c|}
\hline
5 & 3 & 0 & 4 & 2 & -1 \\
\hline
\end{array} \\
& \begin{array}{|c|c|c|c|c|c|c|c|c|c|c|c|}
\hline
2 & -1 & 4 & 1 & 3 & 0 \\
\hline
\end{array} \\
& \begin{array}{|c|c|c|c|c|c|c|c|c|c|c|c|}
\hline
0 & -1 & 3 & -1 & 1 & 4 & 5 & -1 & 2 & -1 & -1 & -1 \\
\hline
\end{array}
\end{align*}

After a value with an id index of 10 ($c_{10}$) is called, the contents of these arrays become:
\begin{align*}
&\begin{array}{|c|c|c|c|c|c|c|c|c|c|c|c|}
\hline
0 & 10 & 8 & 2 & 5 & 6 \\
\hline
\end{array}
\\
&\begin{array}{|c|c|c|c|c|c|c|c|c|c|c|c|}
\hline
12 & 14 & 11 & 9 & 10 & 13\\
\hline
\end{array}
\\
&\begin{array}{|c|c|c|c|c|c|c|c|c|c|c|c|}
\hline
5 & -1 & 0 & 4 & 2 & 1\\
\hline
\end{array}
\\
&\begin{array}{|c|c|c|c|c|c|c|c|c|c|c|c|}
\hline
2 & 5 & 4 & -1 & 3 & 0\\
\hline
\end{array}
\\
&\begin{array}{|c|c|c|c|c|c|c|c|c|c|c|c|}
\hline
0 & -1 & 3 & -1 & -1 & 4 & 5 & -1 & 2 & -1 & 1 & -1\\
\hline
\end{array}
\end{align*}
Notice that only 2 values of each array are changed. The reason this Cache Simulator is independent in runtime of $M$ is because only 1 or 2 values (oldest and youngest) are changed, regardless of the size of $M$. For example, in \verb|nextyounger|, the second value is -1, indicating that $c_{10}$ is now the youngest value. The sixth and final value is changed from -1 to 1, since it is now the second youngest element and points to the youngest element $c_{10}$.
\subsection{Comparing Execution Time}
We will give the data for the amount of time the $\mathcal{O}(M)$ algorithm took to run versus that for the $\mathcal{O}(1)$ algorithm (for the sake of standardization, $b_i=b_j=b_k=1$). Note these values of time were based on a 1.1 GHz Dual-Core Intel Core i3 chip:
\newline

\begin{table}[H]
\centering
\caption{$\mathcal{O}(1)$ vs. $\mathcal{O}(M)$ time, LRU Cache Simulator}
\begin{tabular}{|c|c|c|c|}
\hline
$M$, $n$, values & $\mathcal{O}(1)$ time & $\mathcal{O}(M)$ time & Ratio of $\mathcal{O}(M)$ / $\mathcal{O}(1)$\\

\hline
$M$ = 10, $n$ = 100 & 0.143381 & 0.163240 & 1.13851 \\
$M$ = 100, $n$ = 100 & 0.094647 & 1.170117 & 12.363 \\
$M$ = 1000, $n$ = 100 & 0.100943 & 5.878090 & 58.2318 \\
$M$ = 10000, $n$ = 100 & 0.176576 & 60.644389 & 343.446\\
\hline
\end{tabular}
\end{table}
Although the ratios are not exactly increasing by a factor of 10, the time savings of the $\mathcal{O}(1)$ Simulator are clear. 
\newline
\newline
Now we will give the time taken to run purely the elementary multiplication steps in matrix-matrix multiplication. Note that every standard algorithm for matrix-matrix multiplication, blocked or not, performs roughly $n^3 + (n-1)*(n^2) = 2n^3 - n^2$ elementary multiplications. We use this value to compute flops/sec below.

\begin{table}[H]
\centering
\caption{Varying time to perform computations}
\begin{tabular}{|c|c|c|c|c|c|}
\hline
$n$ & 1x1x1 time & flops/sec & 10x10x10 time & flops/sec & ratio of 10x10x10/1x1x1 time \\

\hline
100 & 0.013484 & 1.48 x $10^8$ & 0.004120 & 4.8 x $10^8$ & 0.306\\
300 & 0.286041 & 1.88 x $10^8$ & 0.113975 & 4.73 x $10^8$ & 0.398\\
500 & 1.979601 & 1.26 x $10^8$ & 0.770256 & 3.24 x $10^8$ & 0.389\\
700 & 4.305378 & 1.59 x $10^8$ & 1.687488 & 4.06 x $10^8$ & 0.392\\
900 & 9.096630 & 1.60 x $10^8$ & 3.408917 & 4.27 x $10^8$ & 0.375\\
\hline
\end{tabular}
\end{table}

What is apparent is that a matrix-matrix multiplication algorithm, with a blocking of 10x10x10, takes roughly a third of the time to run as a matrix-matrix multiplication algorithm with a blocking of 1x1x1. This is a simple illustration of the advantage of blocking.
\newline
\newline
\newline
\section{Conclusion}
We have shown that, given a cache of size $M$, the optimal combination of ($b_i, b_j, b_k$) for matrix-matrix multiplication is ($b$,$b$,1) with
$ b = \lfloor \sqrt{M+1}\rfloor -1$ for communication and execution time both graphically and theoretically. 
We also have demonstrated that the standardly used LRU cache policy can adversely affect the
I/O of an optimal algorithm (and that LFU is worse yet). Based on this observation, 
we provided a way to obtain the optimal amount of I/O in
an LRU cache policy context.
Additionally, we have demonstrated the superiority the $\mathcal{O}(1)$ Cache Simulator possesses in efficiency over the $\mathcal{O}(M)$ Cache Simulator and the computer science behind both algorithms. In the future, we hope to apply the concept of the Cache Simulator to many other algorithms in linear algebra, such as LU Decomposition, QR Factorization, Cholesky Factorization, etc.. We hope that readers will find the Cache Simulator useful in efficient, practical applications of algorithms in conjuction with a cache.   
\bibliographystyle{plain}
\bibliography{neil_biblio}

\begin{thebibliography}{1}

\bibitem{dongarra-2008}
Jack Dongarra, Jean-Fran{\c c}ois Pineau, Yves Robert, Zhiao Shi, and Fr{\'e}d{\'e}ric Vivien.
\newblock Revisiting matrix product on master-worker platforms.
\newblock {\em International Journal of Foundations of Computer Science}, 19(6):1317--1336, 2008.

\bibitem{hong.81.stoc}
Jia-Wei Hong and H.~T. Kung.
\newblock {I/O} complexity: The red-blue pebble game.
\newblock In {\em Proc. of the 13th Annual {ACM} Symposium on Theory of Computing (STOC '81), May 11-13, 1981, Milwaukee, Wisconsin, {USA}}, pages 326--333, 1981.

\bibitem{toledo.jpdc}
Dror Irony, Sivan Toledo, and Alexandre Tiskin.
\newblock Communication lower bounds for distributed-memory matrix multiplication.
\newblock {\em Journal of Parallel and Distributed Computing}, 64(9):1017--1026, 2004.

\bibitem{olpsr:pldi20:20}
Auguste Olivry, Julien Langou, Louis-No{\"e}l Pouchet, P.~Sadayappan, and Fabrice Rastello.
\newblock Automated derivation of parametric data movement lower bounds for affine programs.
\newblock In {\em PLDI 2020: Proceedings of the 41st ACM SIGPLAN Conference on Programming Language Design and Implementation}, page 808–822, June 2020.

\bibitem{sllv:arxiv:19}
Tyler~Michael Smith, Bradley Lowery, Julien Langou, and Robert~A. van~de Geijn.
\newblock A tight {I/O} lower bound for matrix multiplication.
\newblock Technical Report 1702.02017, arXiv, 2019.

\end{thebibliography}

\end{document}